\documentstyle[11pt,paspconf,epsf]{article}
\begin{document}

\vskip -1in
{\tt \noindent To appear in\\ 
\underbar{The Hy-Redshift Universe:\ Galaxy Formation and
Evolution at High Redshift},
eds.\ A.J.\ Bunker \& Wil J.M.\ van Breugel (ASP Conf.\ Series), 1999}

\title{Extremely Red Objects in the Field of QSO~1213--0017:\\
A Galaxy Concentration at $z$ = 1.31}

\author{Michael C. Liu} 
\affil{Department of Astronomy, University of California, Berkeley, CA 94720}

\author{Arjun Dey\altaffilmark{1}}
\affil{National Optical Astronomy Observatories, Tucson, AZ 85719}

\author{James R. Graham}
\affil{Department of Astronomy, University of California,
    Berkeley, CA 94720}

\author{Charles C. Steidel\altaffilmark{2} and Kurt Adelberger}
\affil{Palomar Observatory, Caltech 105-24, Pasadena, CA 91125}

\altaffiltext{1}{Hubble Fellow}
\altaffiltext{2}{Visiting Astronomer, Kitt Peak National Observatories,
National Optical Astronomy Observatories, which is operated by
Associated Universities for Research in Astronomy, Inc., under
cooperative agreement with the National Science Foundation.}

\begin{abstract} 
We have discovered an excess of extremely red objects (EROs) surrounding
the \mbox{$z=2.69$} quasar QSO~1213--0017 (UM~485).  Optical/IR colors
for these galaxies are consistent with $z=1-2$ ellipticals, and there
are at least 5 galaxies with spectroscopic redshifts at
$z\approx1.31$. Keck optical spectra for 3 of the red galaxies show
rest-frame~UV breaks resembling local elliptical galaxies.  Our initial
results suggest a coherent structure in redshift, possibly arising from
a massive galaxy cluster.
\end{abstract}

\keywords{
galaxies: elliptical and lenticular, CD --- 
galaxies: clusters: general --- 
galaxies: evolution --- 
galaxies: stellar content ---
infrared: galaxies}

\section{Introduction}

The nature of extremely red objects (EROs) remains an open question in
understanding the faint galaxy population at $z>1$.  First identified by
Elston et al. (1988), the identity of these objects, defined by their
extreme optical/IR colors ($R-K\ga6$), has proven elusive due to the
optical faintness of the population.  The few EROs with spectroscopic
redshifts form a motley $z\ga1$ collection, including both ultraluminous
dusty star-forming galaxies like HR~10 (Graham \& Dey 1996; Dey et al.\
1999) and passively-evolving luminous ellipticals like LBDS~53W091
(Dunlop et al.\ 1996).  Both of these types of systems are of great
utility in learning about the formation of galaxies and clusters of
galaxies.

We are conducting an on-going study of EROs, using ground-based optical
and near-IR imaging to assemble large samples amenable to statistical
studies and Keck optical/near-IR spectroscopy to determine redshifts and
physical properties. Deciphering the identity of EROs from broad-band
colors alone is ambiguous --- spectroscopy is essential.  In addition,
given the population's apparent heterogeneity, a reasonably large sample
of objects needs to be studied to understand the nature and relative
abundances of the subsets, instead of the spectroscopy of individual
EROs which has been done to date.

We have discovered a concentration of EROs in the field of
QSO~1213--0017, a quasar at $z=2.69$. In this proceedings, we highlight
some of our results. The complete analysis and its details are presented
in Liu et al.\ (1999).

\section{Optical/IR Imaging: An Excess of EROs}

\begin{figure}[t]
\plotfiddle{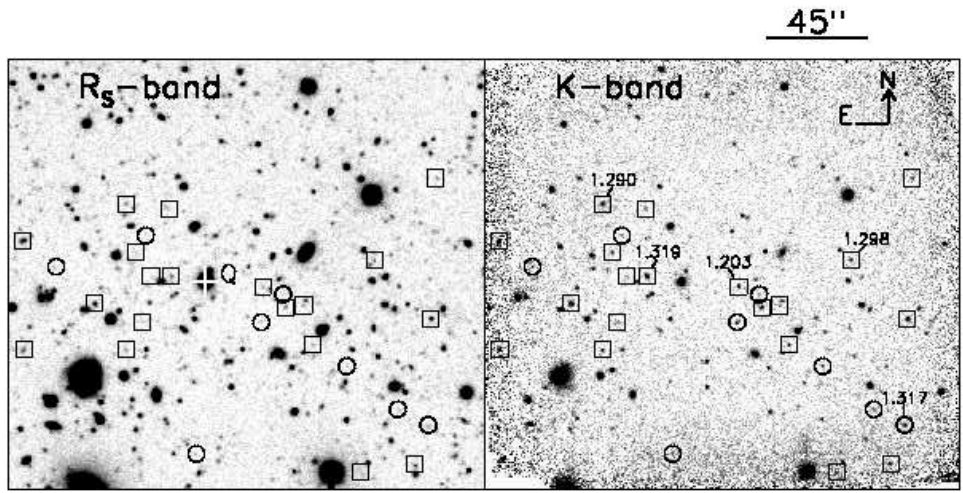}{2.0in}{90}{60}{60}{240}{-80}
\vskip 0.45in
\caption{$R_S$ and $K$-band images of the field. Each is 3\farcm5
$\times$ 3\farcm1. Objects with $R_S-K>6$ are circled and those with
\mbox{$R_S-K=5-6$} are marked with a square.  The $z=2.69$ quasar is
marked with a white cross and a ``Q''. Objects with spectroscopic
redshifts are labelled by their redshift.}
\label{twopanel}
\end{figure}

Figure \ref{twopanel} shows optical and IR images obtained from KPNO of
the 1213--0017 field.  A population of objects with red ($R_S-K>5$)
colors are seen, with the brightest ones being $K\approx18$.  These are
as red as the expected appearance of $z\ga1$ passively evolving
ellipticals.  Most of the sources are resolved in HST $F814W$ imaging so
they are certainly galaxies and not M~stars. We also have $J$ and
$H$-band images for a subset of the galaxies; their optical/IR colors
are also consistent with $z=1-2$ ellipticals (Liu et al.\ 1999).

To gauge the significance of these red galaxies, we compare with field
ERO counts of Thompson et al.\ (1999). After accounting for the
difference between our $R_S$ filter ($\lambda_c\approx6930$~\AA) and the
$R_{CADIS}$ filter of Thompson et al.\ ($\lambda_c\approx6480$~\AA), we find 4
galaxies with $R-K>6$ and $K\le19$ in the deepest 6.2~arcmin$^2$ of our
images. Compared to the 0.24 objects expected from the field counts,
this is a factor of $\approx$15 excess, with a factor of 6 being the
95\% confidence value.

\section{Keck Optical Spectroscopy}

We identified spectroscopic features for 5 of the \hbox{1213--0017} red
galaxies (Table~\ref{table-spectra}). Two of them are emission line
galaxies. Galaxy R6 shows a single narrow emission line which we
identify as [O~II] at $z=1.203$.  The spectrum of R7 reveals the
presence of an active galactic nucleus, showing strong lines of [O~II]
and C~II] with intrinsic widths of $\approx$1800~km~s$^{-1}$.  

\begin{figure}
\plotfiddle{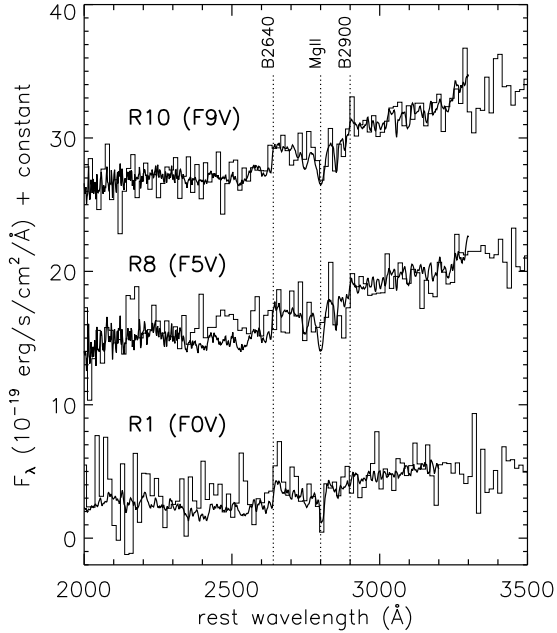}{2.9in}{0}{40}{40}{-150}{-40}
\caption{Spectra of the 3 continuum-break galaxies in the 1213--0017
field, averaged in 7 pixel bins.  The R8 spectrum is offset by +13 along
the y-axis and the R10 spectrum by +26.  Overplotted with the thick
lines are IUE spectra of F~dwarfs, which have been normalized to the
galaxy flux longward of 3000~\AA.  \label{absrestframe}}
\label{fig-2}
\end{figure}

The other 3 red galaxies are devoid of strong emission lines, but they
do show continuum breaks identifiable as the rest-frame mid-UV breaks at
2640~\AA\ and 2900~\AA\ (Figure~\ref{absrestframe}). At least two of the
red galaxies may also have Mg~II in absorption.  The UV break features
are characteristic of the oldest stellar populations, such as Galactic
and M31 globular clusters and the cores of local elliptical galaxies.
They have also been identified in high-redshift ellipticals, e.g.,
LBDS~53W091 at $z=1.55$. For old stellar populations, this spectral
region is expected to be dominated by light from stars near the main
sequence turnoff so, in principal, the strengths of these features can
be used to infer the age of the population --- at least the portion
emitting the bulk of the mid-UV flux --- by basically measuring the
turnoff mass (e.g., Spinrad et al.\ 1997).  
We do not attempt this, but we do point out that real variations exist
in the UV colors and spectral break amplitudes of R1, R8, and R10,
suggesting heterogeneous stellar populations and star formation
histories even though they are at a common location in the Universe and
have similar $K$-band luminosities.

\begin{table}
\caption{Spectroscopic Redshifts in the 1213--0017 Field} \label{table-spectra}
\begin{center} \footnotesize
\begin{tabular}{lcclcl}
Id \# &$K$ (mag) & $R_S-K$ & morphology\tablenotemark{a} & 
$z$ & spectral features\\ 
\tableline
&&&&&\\
R1  & 18.30 & 6.14 & ---           & 1.317 & B2640, MgII, B2900 \\
R6  & 18.89 & 5.69 & diffuse       & 1.203 & [O II] \\
R7  & 18.04 & 5.56 & compact+disk? & 1.319 & [O II], C II], [Ne IV], C III]?\\
R8  & 18.94 & 5.02 & ---           & 1.298 & B2640, MgII?, B2900, D4000? \\
R10 & 18.19 & 5.98 & edge-on disk  & 1.290 & B2640, MgII, B2900, D4000?   \\
&&&&&\\
Mg~II\tablenotemark{b} &    &&  &  1.3196 &\\
                                       &&& &  1.5534 & \\
\end{tabular}
\vskip -0.4in
\end{center}
\tablenotetext{a}{As seen in HST $F814W$ images. Sources with a blank
line ``---'' were not observed.}

\tablenotetext{b}{Mg~II absorption system in the spectrum of
QSO~1213--0017 (Steidel \& Sargent 1992).}
\end{table}

\section{A Coherent Structure at $z$ = 1.31}

We have found a significant excess of EROs in this field, and of the 5
galaxies with spectroscopic redshifts, 4 lie close together.
Moreover, the spectrum of
QSO~1213--0017 shows Mg~II absorption at $z=1.3196$, similar to R7's
redshift.  It is unlikely that R7 is the absorber as it is 15\arcsec\
from the QSO (65~$h^{-1}$~kpc for $\Omega=1, \Lambda=0$), too large
compared to expected sizes of absorbers (Steidel 1993). In addition, our
HST images show some faint galaxies closer to the QSO, making them
better candidates for the absorber.  We therefore have a total of 5
galaxies close in redshift and within $\approx3~h^{-1}$~Mpc on the
sky. Their unweighted mean redshift is 1.309 with a standard deviation
of $1810\pm580$~km~s$^{-1}$ and a full range of 3800~km~s$^{-1}$ in the
mean rest frame.

This is the highest redshift concentration of old, red galaxies
published to date which has been spectroscopically confirmed. Further
data are needed to gauge if this is a genuine massive cluster, both
X-ray images to search for hot intracluster gas in a deep potential
and more redshifts to ascertain the velocity distribution. Finally, we
point out that galaxies R8 and R10 lie at $z=1.290-1.298$ while the
other 3 are at $z=1.317-1.320$, i.e., there is a ``gap'' of
2500~km~s$^{-1}$ in the mean rest frame, though there is no clear
segregation on the sky. These may be two separate physical entities
(e.g., filaments/sheets/sub-clusters), but this speculation lies beyond
the available data.



%
%

\end{document}